\documentclass[amsmath,amssymb,aps,twocolumn,pre,floatfix]{revtex4}

\usepackage{mathrsfs}
\usepackage{amsmath}
\usepackage{amssymb}
\usepackage{color}
\usepackage{graphicx}	
\usepackage{bm}
\usepackage{ulem} 
\usepackage{titlesec}
\titleformat{\paragraph}[runin]
{\bfseries\scshape}{\theparagraph}{1em}{}
\usepackage{braket}
\usepackage{multirow}
\usepackage{xcolor}
\usepackage[printwatermark]{xwatermark}

\newcommand{\be}{\begin{equation}}
\newcommand{\ee}{\end{equation}}
\newcommand{\bef}{\begin{figure}}
\newcommand{\eef}{\end{figure}}
\newcommand{\bea}{\begin{eqnarray}}
\newcommand{\eea}{\end{eqnarray}}

\begin{document}
\title{Computational design of armored nanodroplets as nanocarriers for encapsulation and release under flow conditions}
\author{Fran\c cois Sicard$^{1,2}$}
\thanks{Corresponding author: \texttt{francois.sicard@free.fr}.}
\author{Jhoan Toro-Mendoza$^3$}
\affiliation{$^1$ Department of Physics and Astronomy, University College London, WC1E 6BT London, UK}
\affiliation{$^2$ Department of Chemical Engineering, University College London, WC1E 7JE London, UK}
\affiliation{$^3$Centro de Estudios Interdisciplinarios de la Fisica, Instituto Venezolano de Investigaciones Cientificas, Caracas 1020A, Venezuela}
\begin{abstract}
Nanocarriers are nanosized materials commonly used for targeted-oriented 
delivery of active compounds, including antimicrobials and small-molecular drugs. 
They equally represent fundamental and engineering challenges 
since sophisticated nanocarriers must show adequate structure, stability, 
and function in complex ambients.
Here, we report on the computational design of a distinctive class of nanocarriers, 
built from buckled armored nanodroplets, able to selectively encapsulate or release 
a probe load under specific flow conditions. 
Mesoscopic simulations offer detailed insight into the interplay 
between the characteristics of laden surface coverage and 
evolution of the droplet morphology. 
First, we describe in detail the  formation of \textit{pocket-like} structures 
in Pickering emulsion nanodroplets and their stability under 
external flow. Then we use that knowledge to test the capacity of these 
emulsion-based pockets to yield flow-assisted encapsulation or expulsion 
of a probe load. Finally, the rheological properties of our model carrier 
are put into perspective with those of delivery systems employed in pharmaceutical 
and cosmetic technology.
\end{abstract}
%
\keywords{Nanocarriers, Pickering emulsion, buckling, shear flow, encapsulation, Dissipative Particle Dynamics}

\maketitle

%
Over the last decade, special attention has been assigned 
to the design, characterization, and development of nanocarrier 
systems, which can have potential in targeted-oriented active molecule 
delivery. They offer remarkable advantages in a wide range 
of industrial and medical applications, including food~\cite{2012-FBT-Silva-Vicente}, cosmetic~\cite{2019-FC-Aziz-Ibrahim} 
and pharmaceutical industries~\cite{2018-NC-Rosenblum-Peer}.
In this context, nanoparticle-stabilized emulsions, aka Pickering emulsions~\cite{Pickering1907}, have been intensively used as 
drug-delivery vehicles in topical medication~\cite{2009-IJP-Frelichowska-Chevalier}, 
where their surfactant-free character makes them attractive for 
different applications since surfactants often produce adverse effects, 
such as irritation and haemolytic disturbances~\cite{2005-JPBA-Aparicio-Mitjans,2013-CSAPES-Chevalier-Bolzinger}.
They can also serve as ideal compartments for reactions catalyzed by 
nanoparticles (NPs) attached at the oil-water 
interfaces~\cite{2014-ACSCatal-Shi-Resasco,2015-ACSCatal-Faria-Resasco,2017-PPSC-Qu-He} and 
can be used in bacterial recognition technologies.~\cite{2014-ACIE-Shen-Ye,2019-SciRep-Horvath-Szechenyi} 
Another important and useful advantage of Pickering emulsions 
over conventional surfactant-stabilized systems is their enhanced 
stabilization against coalescence~\cite{2016-FD-Sicard-Striolo} 
and their smaller environmental footprint~\cite{2020-Engineering-Gonzalez-Miele}. 
While tremendous progress has been made in particle-based microfluidic 
technology~\cite{2012-RPP-Seeman-Herminghaus,2019-CSC-Chacon-Baret}, the characteristics of Pickering emulsions pose a number 
of intriguing physical questions, including a thorough 
understanding of the perennial lack of detail about how particles arrange 
at the liquid/liquid interface. Predicting and controlling 
this interfacial arrangement is even more challenging under flow conditions.\\

Here, we report on the computational design of a new class 
of nanocarriers built from Pickering nano-emulsions, which exhibit 
a \textit{pocket-like} morphology able to encapsulate or release a probe 
load under specific flow conditions.
Dissipative Particle Dynamics (DPD) is employed as a 
mesoscopic simulation method~\cite{1997-JCP-Groot-Warren} 
with two aims: (1) to describe in detail the formation of 
\textit{pocket-like} structures in Pickering nanodroplets 
and their stability under specific 
flow conditions and then (2)  to test the capacity 
of the formed pockets to encapsulate or release a probe load.
Also, the physical properties of our model carrier 
are put into perspective within the conditions encountered in 
the high-shear regime of spreading topical medication on 
the skin and the transport of targeted carriers  in pathological 
alterations of the vascular system.
Despite technological advances in experimental methods 
to control NP assembly at fluid interfaces~\cite{2012-Nanoscale-Reguera-Moglianetti,2016-Nanoscale-GinerCasares-Reguera}, 
the inherent limitation in experimental resolution eludes direct access 
to local observables, such as the particles’ three-phase contact angle 
distribution and the details of the particles' interfacial network \cite{2016-SoftMatter-Binks-Yin} 
presenting complex geometries, while
these pieces of information can be accessed by numerical 
simulations~\cite{2016-FD-Sicard-Striolo,2017-Nanoscale-Sicard-Striolo,SICARD2018167,2019-ACSNano-Sicard-Striolo}.\\

\begin{figure*}[t]
\includegraphics[width=0.95 \textwidth, angle=-0]{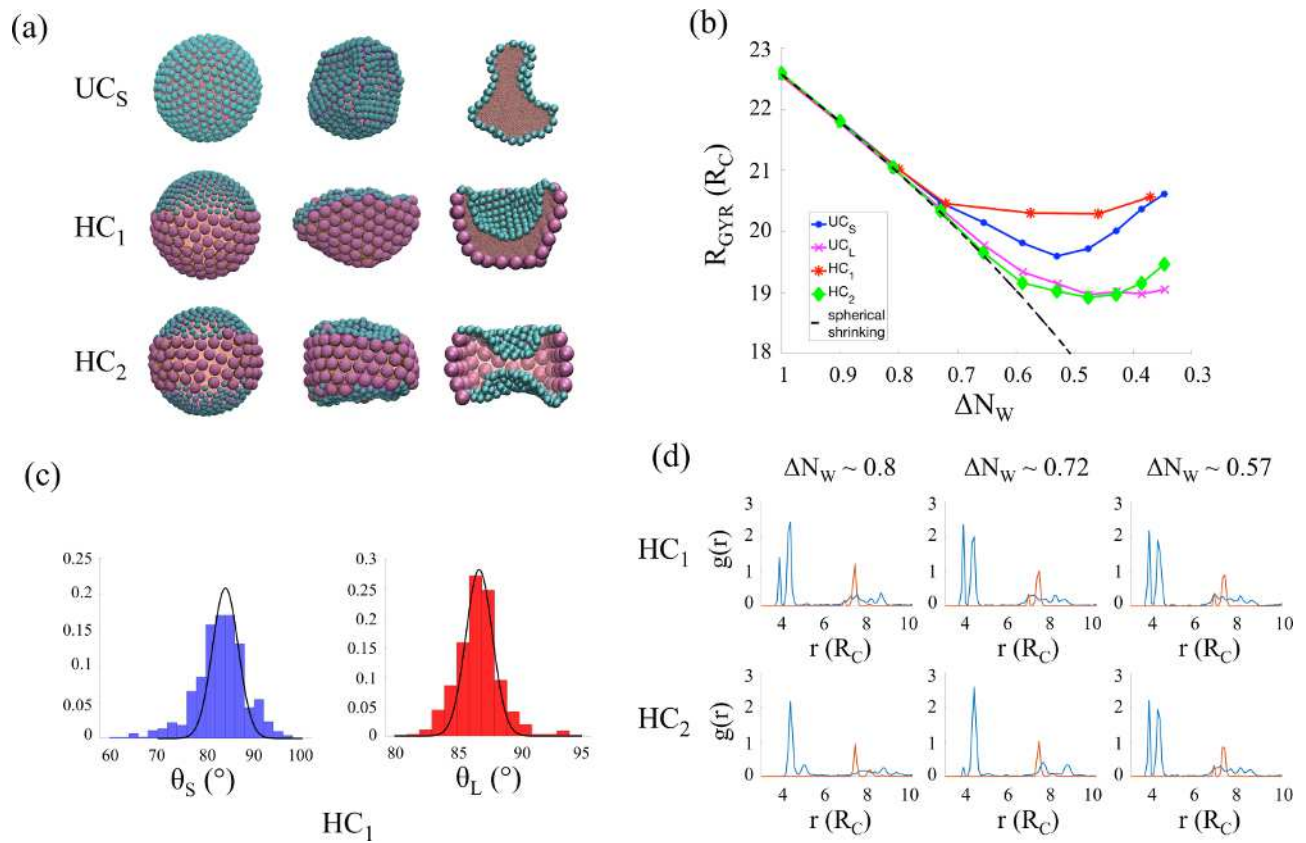}
 \caption{\textbf{Formation of pocket-like structures.} 
 (a) Simulation snapshots representing the initial (left) 
 and final (center) water in oil droplets armored with different 
 nanoparticles surface coverages, obtained fom the evaporation process: 
 uniformly covered droplet with small NPs ($\textrm{UC}_L$)
 and heterogeneously covered droplets with either each hemisphere 
 covered with small or large NPs ($\textrm{HC}_1$) or three distinct 
 layers made of small-large-small NPs ($\textrm{HC}_2$). 
 The cross-sectional view of each system is also shown (right). 
 Cyan and purple spheres represent the small and large Janus NPs, 
 respectively. The detailed structure of the NPs is shown in Fig.~S1a 
 in the SI. Pink spheres represent water beads. 
 The oil molecules surrounding the system are not shown for clarity. 
 (b) Evolution of the radius of gyration, $R_{\textrm{GYR}}$, 
 of $\textrm{UC}_S$, $\textrm{UC}_L$, $\textrm{HC}_1$, and $\textrm{HC}_2$, 
 as a function of the dimensionless parameter $\Delta N_W = N_W/N_W^{(0)}$. 
 $N_W$ represent the number of water beads that remain in the droplet 
 after each removal and $N_W^{(0)}$ is the initial number of water beads. 
 The statistical errors are estimated as one standard deviation 
 from the average obtained for equilibrated trajectories and 
 are always smaller than the symbols. 
 The dashed lines represents the spherical-shrinking regime 
 defined as $R_{\textrm{GYR}} \sim (\Delta N_W)^{1/3}$. 
 (c) Three-phase contact angle distribution of small (blue) and 
 large (red) NPs for $\textrm{HC}_1$ after the last pumping/equilibration 
 iteration ($\Delta N_W \sim 0.35$).
 (d) Evolution of the radial distribution function, 
 $g(r)$, with $r$ the distance between the center of the NPs, 
 of small (blue) and large (red) NPs for $\textrm{HC}_1$ (b) 
 and $\textrm{HC}_2$ (c).
 }
\label{fig1}
\end{figure*}

Water nanodroplets coated with spherical NPs with 
different diameters and immersed in an organic solvent are considered. 
The coating is formed by Janus NPs (particles whose surface shows two 
distinct wetting properties)~\cite{2017-Langmuir-Zhang-Granick,2019-ACSANM-Agrawal-Agrawal} 
whose initial three-phase contact angles result in maximum adsorption energy at the fluid-fluid interface~\cite{2000-Langmuir-Binks-Lumsdon}.
Hence, we are able to quantify the role played by homogeneous and heterogenous NP surface 
coverage at the emulsion droplet interface when the volume of the droplet 
is reduced. In particular, we observe in detail the formation of 
crater-like depressions with selective geometry, 
which can structurally favour the loading of a probe load.
The flow conditions clearly affect the dynamical response of the \textit{pocket-like} 
armored nanodroplets. Under specific conditions, we observe
the formation of long-lived anisotropic structures, characteristic 
of a jammed particle coverage at the liquid-liquid interface. 
Furthermore, we examine the capacity of the system to control the flow-assisted encapsulation 
or release of a probe load, which depends on the interplay between 
NP surface coverage, the level of buckling, and the shear flow conditions.\\

%
\textbf{System characteristics.} In Fig.~\ref{fig1}a we show 
representative snapshots of 
water emulsion nanodroplets in organic solvent (decane) 
stabilized with Janus NPs. The scaled temperature 
in the DPD framework is equivalent to $298.73$ K. 
The details of the numerical parametrization and NP structures are given 
in the Methods section and the Supporting Information (SI).
The configurations differ by the size of the NPs and 
the characteristics of the surface coverage. We consider 
small (S) and large (L) NPs with diameters $d_S \sim 2.2$ nm 
and $d_L \sim 4.5$ nm, whose diffusion coefficients measured 
on a planar decane/water interface are  
$D_S \sim 4.7 \pm 3.1 \times 10^{-7}$ $\textrm{cm}^2~\textrm{s}^{-1}$ 
and 
$D_L \sim 1.8 \pm 0.7 \times 10^{-7}$ $\textrm{cm}^2~\textrm{s}^{-1}$,
respectively (see Methods and Fig.~S1c in the SI). 
The NPs are originally 
located at the surface of the emulsion nanodroplets of 
diameter $d_D \sim 45$ nm.
Similar NP surface coverage $\phi \sim 0.8$, as defined 
in Ref~\cite{2013-Langmuir-Luu-Striolo,2017-Nanoscale-Sicard-Striolo}, is considered on the armored nanodroplets. 
This yields similar initial three-phase contact angles 
$\theta_S \sim 84.1^{\circ} \pm 2.7^{\circ}$ and 
$\theta_L \sim 86.8^{\circ} \pm 1.1^{\circ}$ for the small and large NPs, 
respectively (see Methods and Fig.~S1b in the SI), in qualitative 
agreement with simulations~\cite{2011-Langmuir-Fan-Striolo,2012-PRE-Fan-Striolo,2016-FD-Sicard-Striolo} and experimental observations~\cite{2010-PCCP-Arnaudov-Paunov}.
From the error bars estimated, it is observed that the small NPs 
are more sensitive to thermal fluctuations at the interface compared
to the large ones, characteristic of the increase of the adsorption
energy with the particle radius~\cite{2001-Langmuir-Binks-Fletcher,2007-JCP-Jiang-Granick,2020-PCCP-Khedr-Striolo}. 
We also measure the decrease of the interfacial tension, 
$\Delta \gamma_S$ and $\Delta \gamma_L$, for small and large NPs 
at planar interfaces for similar NP surface coverage (see Methods). 
We obtain
$\Delta \gamma_S = \gamma_0-\gamma_S \sim 5.1~\textrm{mN}.\textrm{m}^{-1}$
and 
$\Delta \gamma_B = \gamma_0-\gamma_B \sim 2.2~\textrm{mN}.\textrm{m}^{-1}$, 
with $\gamma_0 \sim 51.7~\textrm{mN.m}^{-1}$ the interfacial tension for a planar decane/water 
interface~\cite{2012-PRE-Fan-Striolo,2019-ACSNano-Sicard-Striolo}, 
and $\gamma_{S,B}$ the interfacial tension when the interface 
is covered with small or large NPs, respectively.  
In particular, large NPs have less effect on the 
reduction of the interfacial tension and are less diffusive than smaller ones, in qualitative agreement 
with simulations~\cite{2020-PCCP-Khedr-Striolo} and experimental observations~\cite{2011-Small-Wang-Koynov}. A lower mobility along with the size of the NPs, will play a key role in the pocket formation.\\

\textbf{Formation of pocket-like structures.} The volume of the 
droplets is systematically reduced, 
by iteratively pumping a small constant proportion of water 
molecules out of the droplets and letting the systems equilibrate 
between each iteration, until the systems present dimples 
and cups at the droplet interface followed by the formation 
of crater-like depressions, characteristic 
of the buckling instability~\cite{2010-Langmuir-Datta-Weitz,2017-Nanoscale-Sicard-Striolo,2018-SoftMatter-Gu-Botto} (see details in the SI). 
This process is physically equivalent to a process of solubilization of the dispersed phase into the solvent~\cite{2010-Langmuir-Datta-Weitz,2017-Nanoscale-Sicard-Striolo}.
We arbitrarily stop the pumping 
when the number of water molecule constituting 
the droplets reaches the value 
$\Delta N_W = N_W/N_W^{(0)} \sim 0.35$,  
where $N_W^{(0)}$ and $N_W$ are the initial number of water beads 
and the number of water beads remaining in 
the droplets, respectively.

\begin{figure*}[t]
\includegraphics[width=0.9 \textwidth, angle=-0]{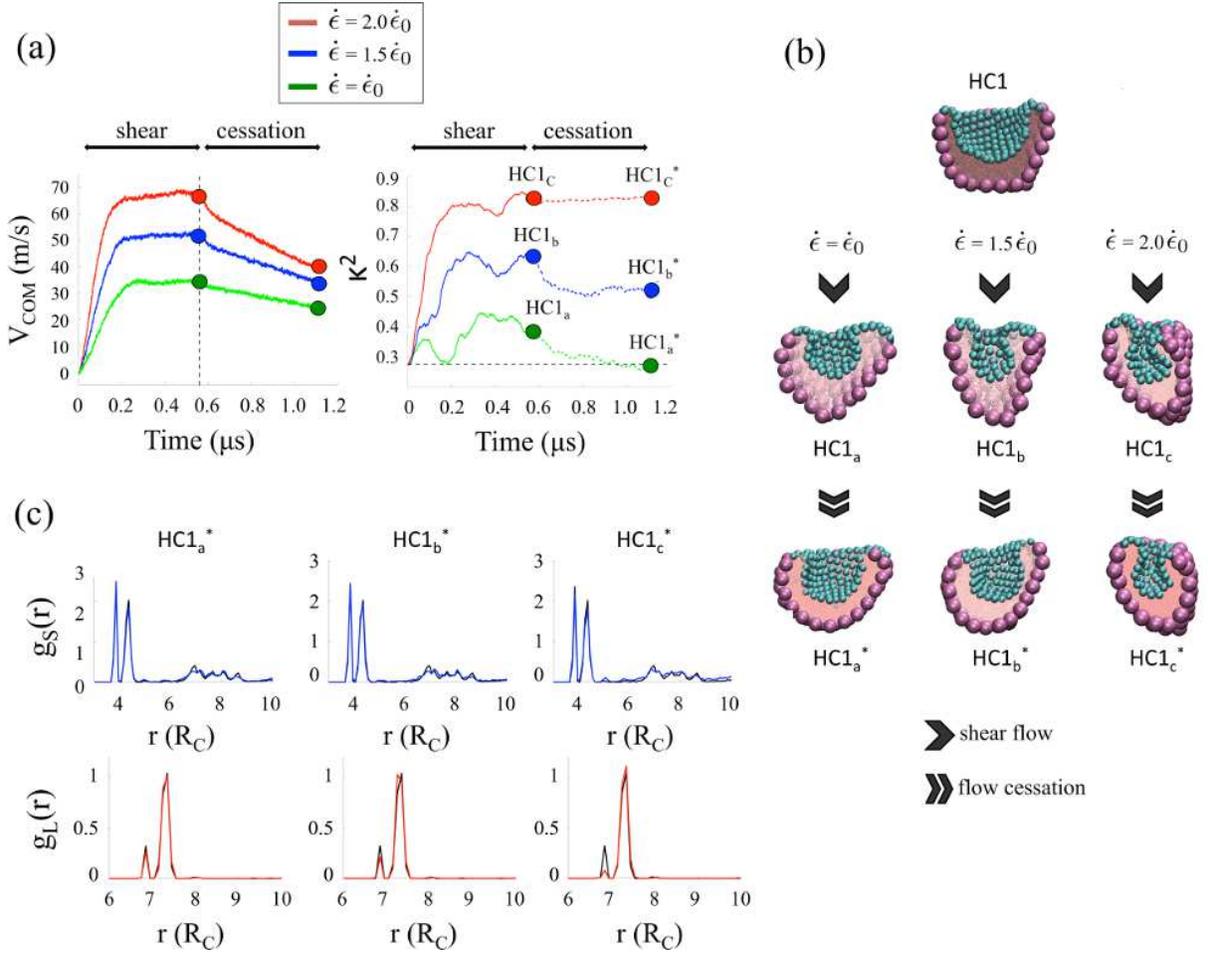}
\caption{\textbf{Dynamical response under shear flow.} 
(a) Temporal evolution of the velocity center of mass $V_{COM}$ 
 and the relative shape anisotropy $\kappa^2$ of $\textrm{HC}_1$ 
 subjected to shear flow and after abrupt shear cessation for three 
 different values of the shear rate $\dot{\epsilon}$. 
 The shear flow is continuously applied for a time duration 
 $\Delta t \sim 0.6~\mu s$ before it is abruptly stopped and 
 the structure relaxes for another $\Delta t \sim 0.6~\mu s$.
 (b) Representative snapshots of the armored nanodroplets obtained 
 just before the flow cessation ($ t \sim 0.6~\mu s$) and at the end 
 of the simulation ($t \sim 1.2~\mu s$) are shown. 
 Cyan and purple spheres represent the small and large Janus NPs, 
 respectively. The detailed structure of the NPs is shown in Fig.~S1a 
 in the SI. Pink spheres represent water beads. 
 The oil molecules surrounding the system are not shown for clarity. 
 (c) Radial distribution function, $g_S(r)$ and $g_L(r)$, 
 with $r$ the distance between the center of the NPs, 
 of small (blue) and large (red) NPs for $\textrm{HC1}_{a,b,c}^*$. 
 The corresponding radial distribution functions measured 
 before the shear flow is applied ($\textrm{HC1}$) 
 are shown in black color for comparison.
 }
\label{fig2}
\end{figure*}
In Fig.~\ref{fig1}b, we show the evolution of the radius of gyration 
of the emulsion nanodroplets, $R_{\textrm{GYR}}$, as a function of 
the dimensionless parameter $\Delta N_W$. We initially consider 
spherical droplets whose surface is either uniformly covered (UC) 
with NPs of identical diameter or heterogeneously covered (HC) 
with NPs of different diameters. 
In particular, $\textrm{UC}_S$ (respectively $\textrm{UC}_L$) is solely 
covered with small (respectively large) NPs, as shown in  Fig.~\ref{fig1}a.
$\textrm{HC}_1$ and $\textrm{HC}_2$ have each hemisphere covered with small 
and large NPs, or three distinct layers made of 
small-large-small NPs, respectively (cf. Fig.~\ref{fig1}a).
When $\Delta N_W>0.75$, the radii of gyration of four systems follow 
similar evolution, regardless the NP coverage (UC or HC), 
characteristic of a spherical-shrinking regime, 
$R_{\textrm{GYR}} \sim (\Delta N_W)^{1/3}$ (dashed line in Fig.~\ref{fig1}b).
When $\Delta N_W<0.75$, the systems follow different transitions 
from spherical shrinking to buckling, 
depending on the characteristics of the NP interfacial packing originating 
from the difference in surface coverage~\cite{2006-Langmuir-Basavaraj-Vermant}. 
This transition happens when the NP monolayer becomes close 
to its maximum packing, as observed with the evolution of the radial 
distribution function $g(r)$, with $r$ the distance between the center 
of the NPs, shown in Fig.~\ref{fig1}d and Figs.~S2 in the SI.
In particular, $\textrm{UC}_S$ and $\textrm{UC}_L$ show different 
morphological evolutions when $\Delta N_W$ decreases, with $\textrm{UC}_S$ 
entering the buckling regime at larger $\Delta N_W$ than $UC_L$, 
in qualitative agreement with the numerical work of 
Gu et al.~\cite{2018-SoftMatter-Gu-Botto}.
Finally, below $\Delta N_W \sim 0.45$. $R_{\textrm{GYR}}$ 
increases as the droplets can be described as half-sphered.\\

The structures of the armored nanodroplets obtained after 
the last pumping/equilibration iteration are shown 
in Fig.~\ref{fig1}a (central panel). Visual inspection shows different 
folding morphologies, depending on the characteristics 
of the NP coverage. 
Unlike UC where crater-like depressions  
form evenly at the interface when the droplet is subject to a
compressive surface stress, we observe the formation of  
well-localised crater-like depressions when the droplet is 
heterogeneously covered ($\textrm{HC}_1$ or $\textrm{HC}_2$), 
depending on the localisation 
of the interfacial areas covered with small or large NPs. 
Notably, we observe the crater-like depressions form 
in the interfacial areas covered with the smallest NPs, 
where maximum packing of the interfacial network is achieved 
quicker and the interfacial tension is lower than 
those measured for larger NPs.\\ 

The properties of the interfacial layers 
are quantitatively assess via the analysis of the distribution of the three phase contact angles, 
$\theta_C^{(S)}$ and $\theta_C^{(L)}$, of small and large NPs, 
respectively. 
As shown in Fig.~S1b in the SI, $\theta_C^{(S)}$ and $\theta_C^{(L)}$ follow Gaussian distributions in the initial configurations 
($\Delta N_W \sim 1$), where the shape of the droplets is spherical. 
When the volume of $\textrm{UC}_S$ and $\textrm{UC}_L$ is reduced, 
$\theta_C^{(S)}$ and $\theta_C^{(L)}$ uniformly evolve 
from Gaussian to skewed unimodal distributions, 
in line with previous work~\cite{2017-Nanoscale-Sicard-Striolo}. 
The values of the respective 
means, $\mu_S$ and $\mu_L$, and standard deviations, 
$\sigma_S$ and $\sigma_L$, for small and large NPs, respectively, 
are shown in Table~\ref{Tab-thetaC}. 
Whereas the contact angle distributions show a single peak centered 
at the same value as the one measured for the initial configuration, 
$\sigma_S$ and $\sigma_L$ show significant variations 
when the volume of the droplets is reduced, characteristic of  
the skewness of the distribution and the decrease 
of the NP–NP distance (cf. Fig.~S2 in the SI).
\begin{table}[t]
\begin{center}
\begin{tabular}{l c c c c c}
  \hline
  $\Delta N_W$ & {} & $\textrm{UC}_S$ & $\textrm{UC}_L$ & $\textrm{HC}_1$ & $\textrm{HC}_2$\\
  \hline
  \multirow{2}{*}{1.0} & \multicolumn{1}{l}{(S)} & \multicolumn{1}{c}{$84.1^\circ \pm 2.7^\circ$} & \multicolumn{1}{c}{$-$} & \multicolumn{1}{c}{$84.1^\circ \pm 2.7^\circ$} & \multicolumn{1}{c}{$84.1^\circ \pm 2.7^\circ$}\\
                       & \multicolumn{1}{c}{(L)} & \multicolumn{1}{c}{$-$} & \multicolumn{1}{c}{$86.8 \pm 1.1$} & \multicolumn{1}{c}{$86.8^\circ \pm 1.1^\circ$} & \multicolumn{1}{c}{$86.8^\circ \pm 1.1^\circ$}\\
  \hline                     
  \multirow{2}{*}{0.35} & \multicolumn{1}{c}{(S)} & \multicolumn{1}{c}{$82.9^\circ \pm 5.9^\circ$} & \multicolumn{1}{c}{$-$} & \multicolumn{1}{l}{$82.8^\circ \pm 6.0^\circ$} & \multicolumn{1}{c}{$82.4 \pm 6.4$}\\
                      & \multicolumn{1}{c}{(L)} & \multicolumn{1}{c}{$-$} & \multicolumn{1}{c}{$83.6^\circ \pm 9.9^\circ$} & \multicolumn{1}{c}{$86.7^\circ \pm 1.9^\circ$} & \multicolumn{1}{c}{$87.0 \pm 1.8$}\\
  \hline
\end{tabular}
\caption{Measure of the mean ($\mu$) and standard error ($\sigma$) 
of the three phase contact angle distribution in UC and HC droplets 
in the initial ($\Delta N_W \sim 1.0$) and final ($\Delta N_W \sim 0.35$) 
configurations.}
\label{Tab-thetaC}
\end{center}
\end{table}
When the volume of $\textrm{HC}_1$ and $\textrm{HC}_2$ is reduced, 
on the other hand, we observe significant differences in the 
evolution of the distributions of $\theta_C^{(S)}$ 
and $\theta_C^{(L)}$, due to the heterogeneity in NP size and 
surface coverage. 
In particular, the distribution of $\theta_C^{(L)}$ is 
similar to the one measured in the initial configuration, 
while the distributions of $\theta_C^{(S)}$ shows large variability, 
similar to the one measured in $\textrm{UC}_S$, during the buckling 
transition, originating from the difference in packing of the 
monolayer at the droplet interface, as shown in Fig.~\ref{fig1}c. \\

\textbf{Dynamical response under shear flow.} Thereafter, 
we investigate the structural response 
of the buckled armored nanodroplets subjected to shear flow of the surrounding fluid 
using the SLLOD algorithm~\cite{1984-CPR-Evans-Morriss,1984-PRA-EvansMorris} 
coupled with Lee-Edwards periodic boundary 
conditions~\cite{1972-JPC-Lees-Edwards} (see Methods). 
We focus our analysis on 
$\textrm{HC}_1$ whose structural morphology is more likely to yield 
better loading of a probe load (cf. Fig.~\ref{fig1}a).
The minimum value for the shear rate, 
$\dot{\epsilon}_0 \sim 0.9~\textrm{ns}^{-1}$, is set to the one 
for which the initial structure starts showing 
significant deformations. 
The system is first stressed under a constant shear rate, $\dot{\epsilon} = \alpha \times \dot{\epsilon}_0$, 
along the $x$-axis for a time duration $ \Delta t \sim 0.6~\mu\textrm{s}$, 
with the parameter $\alpha=1.0$, $1.5$, and $2.0$. 
The length of the simulation is chosen sufficiently long for 
the velocity center of mass of the droplet, 
$V_{COM}$, to level off to a plateau whose value matches the one 
obtained from stationary velocity profile of laminar flow, 
$V_{COM} = \dot{\epsilon} \times L_y/2$, with $L_y \sim 77$ nm 
the size of the simulation box along the $y$-direction (cf. Fig.~\ref{fig2}a). 
The flow is then abruptly halted and the dynamical stability of 
the nanodroplet  is studied for a time duration 
$ \Delta t\sim 0.6~\mu\textrm{s}$.

\begin{figure*}[t]
\includegraphics[width=0.9 \textwidth, angle=-0]{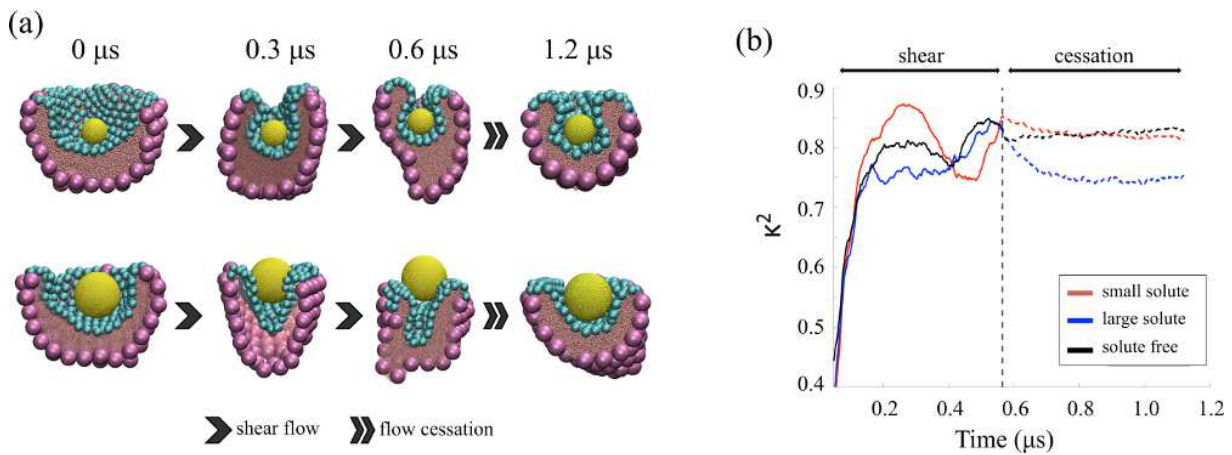}
 \caption{\textbf{Encapsulation and release of load probe.} 
 (a) Representative snapshots of the structural morphology 
 of $\textrm{HC}1$ preliminary loaded with small (S) and large (L) 
 hydrophobic spherical solute of diameter $d_S \sim 7.7$ nm (top panel) 
 and $d_L \sim 15.4$ nm (bottom panel), respectively. 
 Left, middle, and right panels correspond to the initial structure, 
 and those obtained after $0.2~\mu$s and $0.5~\mu$s, respectively. 
 Cyan and purple spheres represent the small and large Janus NPs, 
 respectively. The detailed structure of the NPs is shown in Fig.~S1a 
 in the SI. Pink and gold spheres represent water and solute beads. 
 The oil molecules surrounding the system are not shown for clarity. 
 (b) Representative temporal evolution of the relative shape 
 anisotropy, $\kappa^2$, of $\textrm{HC}_1$ loaded with small and large 
 solute, subjected to shear flow ($\dot{\epsilon} = 2\times \dot{\epsilon}_0$) 
 and after abrupt shear cessation. 
 The shear flow is continuously applied for a time duration 
 $\Delta t \sim 0.6~\mu s$ before it is abruptly stopped and 
 the structure relaxes for another $\Delta t \sim 0.6~\mu s$. 
 The evolution of $\kappa^2$ for the free system is shown for 
 comparison.
 }
\label{fig3}
\end{figure*}
Representative snapshots of the structural 
morphology of the armored nanodroplets obtained after 
$t \sim 0.6~\mu\textrm{s}$ and $t \sim 1.2~\mu\textrm{s}$, 
identified in Fig.~\ref{fig2}a, are shown in Fig.~\ref{fig2}b. 
Visual inspection shows different morphologies depending on 
the intensity of the shear rate and the relaxation of the 
system.
The changes in structural morphology is quantitatively 
assessed with the measure of the relative shape anisotropy 
parameter, $\kappa^2$, which reflects both the symmetry 
and dimensionality of the system~\cite{2011-JPCA-Vymetal-Vondrasek,2013-JCP-Arkin-Janke} 
(see Methods). 
As shown in Fig.~\ref{fig2}a (right panel), we observe the increase of 
$\kappa^2$ at relatively short time until it levels off 
to a plateau when the velocity profile of the laminar fluid 
becomes stationary, and whose value depends on the intensity 
of the shear rate. 
In Figs.~\ref{fig2}b and Fig.~S3a in the SI, 
we observe the increase of 
$\kappa^2$, associated with the elongation of the droplet 
along the deformation axis $x$ and with the squeezing of the 
crater-like depression along the orthogonal $z$-direction.
($\textrm{HC}1_{a,b,c}$).\\

When the flow is abruptly halted at $t \sim 0.6~\mu\textrm{s}$, 
we observe either the relaxation of $\kappa^2$ towards its initial 
value ($\textrm{HC}1_{a}^*$) or the formation of long-lived anisotropic structure ($\textrm{HC}1_{b,c}^*$), depending 
on the intensity of $\dot{\epsilon}$. 
The specificity of the structural morphology of $\textrm{HC}1_{b,c}^*$ 
can be explained by the formation of a jammed particle layer at 
the droplet interface, in qualitative agreement with recently reported experimental 
observations~\cite{2020-SoftMatter-Kaganyuk-Mohraz}. 
To do so, we assess the characteristics of the NP interfacial layer of 
$\textrm{HC}1_{a,b,c}^*$ with the analysis of the three-phase 
contact angle distribution and the NP radial distribution function  
of small and large NPs. 
Within the range of shear rates considered in this work, 
$\theta_C^{(L)}$ follows a Gaussian distribution of 
mean $\mu_L\sim 87.2^{\circ}$ and standard deviation 
$\sigma_L\sim 1.8^{\circ}$, similar the the one measured 
in both the initial and buckled configurations (cf. Fig.~\ref{fig1}c).
$\theta_C^{(S)}$, on the other hand,  shows a  skewed unimodal 
distribution with a central  peak located at  the  same  value  
as  the one  measured  for  both the  initial and buckled 
configurations. The skweness of the distribution does not depend 
significantly on the intensity of the shear rate within 
the standard errors (cf. Fig.~S3 in the SI).

Most importantly, the radial distribution functions, $g_S$ 
and $g_L$, of small and large NPs, respectively, show 
different behaviours depending on the size of the NPs, 
as shown in Fig.~\ref{fig2}c.
Whereas $g_S$ follows the same evolution as the one measured 
in $\textrm{HC}_1$ before the shear rate is applied, 
the evolution of $g_L$ reflects the local reorganisation of 
the layer made solely of large NPs at the droplet interface, 
as shown with the gradual decrease 
of its first peak associated with the first coordination sphere, 
eventually recovering the distribution observed in the initial 
spherical configuration shown in Fig.~\ref{fig1}d.\\

\textbf{Encapsulation and release of probe load.} Our results so far 
allow us to address our second aim of investigating the dynamical response 
of the system under shear stress, when the buckled armored nanodroplet 
is preliminary loaded with a probe load, as shown in Fig.~\ref{fig3}a. 
Then, we determine the ability of $\textrm{HC}1_c$ to lead to 
the encapsulation or release of the solutes under flow conditions 
identical to those studied in the free configuration. 
We consider the largest shear rate, 
$\dot{\epsilon} = 2 \times \dot{\epsilon}_0 \sim 1.8~\textrm{ns}^{-1}$, 
which shows the strongest structural deformation of the system, 
as shown in Figs.~\ref{fig2}a (right panel).
One small ($S_S$) and one large ($S_L$) spherical hydrophobic 
solutes are considered, with radius $r_S^{(s)} \sim 4$ nm 
and $r_L^{(s)} \sim 8$ nm, respectively. The size of $S_S$ and 
$S_L$ is specifically chosen so that they can be preliminary loaded 
in the crater-like depression formed at the interface of 
$\textrm{HC}1$, obtained after the last removal of water 
(cf Fig.~\ref{fig1}a). $S_S$ and $S_L$, however, 
differ in their ability to eventually fit or not in $\textrm{HC}1_c$ 
when the shear stress is applied.
The characteristics of the spherical solutes in the DPD framework 
are given in the Methods section.

The system is first stressed under constant shear rate 
along the $x$-axis for a time duration $\Delta t \sim 0.6~\mu$s, sufficiently long to observe the flow-assisted 
encapsulation or release of the small and large solutes, respectively. 
The flow is then abruptly halted and the relaxation of the system is 
studied for a time duration $\Delta t \sim 0.6~\mu$s. 
In Fig.~\ref{fig3}a we show representative snapshots of the 
systems loaded with the two spherical solutes, 
$S_S$ and $S_L$, at different simulation stages. When the solute is sufficiently small, the 
particle-laden interface folds inward under surface stress  
leading to the encapsulation of the solute.
When the solute is sufficiently large, however, the crater-like 
depression cannot accommodate the solute  
when the system is stressed. Therefore, $S_L$ is progressively expelled 
from the pocket following the narrowing and elongation of 
the nanodroplet. 
As the flow is abruptly halted, the armored nanodroplet relaxes 
its structural morphology, accommodating the solute load 
inside the residual pocket, regardless the size of the load. 

The evolution of the structural morphology of the loaded nanodroplets 
is quantitatively assessed with the estimation of the relative shape 
anisotropy, $\kappa^2$, as shown in Fig.~\ref{fig3}b. 
In particular, we compare the average value of $\kappa^2$ 
in the stationary regime, 
\textit{i.e.} $0.2~\mu s \leq t \leq 0.6~\mu s$, defined as  
$\langle \kappa^2 \rangle = \frac{1}{\Delta t} \int \kappa^2(t) dt$, 
along with the relative change 
$ \delta \kappa^2 = \Big\vert \kappa^2(t=1.2~\mu s) - \kappa^2(t=0.6~\mu s) \Big\vert /\kappa^2(t=0.6~\mu s)$, measured between the beginning ($t=0.6~\mu s$) 
and the end ($t=1.2~\mu s$) of the relaxation period.
\begin{table}[t]
\begin{center}
\begin{tabular}{l c c c }
  \hline \hline
  {} & free & $S_S$ & $S_L$ \\
  \hline
  $\langle \kappa^2 \rangle$ & $0.81 \pm 0.02$ & $0.79 \pm 0.04$ & $0.80 \pm 0.05$ \\
  $ \delta \kappa^2$ & $1.2 \% \pm 0.3\%$ & $4.4 \% \pm 1.6 \%$ & $6.4 \% \pm 4.5 \%$ \\
  \hline \hline
\end{tabular}
\caption{Estimation of the average value of $\kappa^2$ when the system 
reaches a stationary state under flow conditions, $\langle \kappa^2 \rangle$, 
and the relative change $ \delta \kappa^2$ between the beginning 
and the end of the relaxation period. 
Uncertainties are determined by considering three replica of 
the systems, and calculating the standard error.
}
\label{Tab-load}
\end{center}
\end{table}
As shown in Tab.~\ref{Tab-load}, the values of $\langle \kappa^2 \rangle$ 
estimated in the free and loaded configurations do not differ significantly 
within the standard errors, suggesting the pocket-like nanodroplet 
passively encapsulates or expels the small and large solutes, 
respectively, under the flow conditions and solute characteristics 
considered in this work. 
When the flow is abruptly halted, on the other hand, we observe the 
relaxation of the system, which accommodates the solute load
inside the residual pockets. During this process, the relaxation of 
the structural morphology of the loaded nanodroplets differs from 
the solute-free configuration, as quantified with the relative change 
$\delta \kappa^2$ in Tab.~\ref{Tab-load}, in qualitative agreement 
with the visual inspection in Fig.~\ref{fig3}a.\\

\textbf{Perspectives in delivery technology.} 
The flow-assisted encapsulation and release of load probes in 
armored nanodroplets reported so far can be extended  to systems of larger 
dimensions under conditions similar 
to those expected in the high-shear regime of spreading topical 
medication on the skin  (such as creams and ointments) 
and the transport of targeted carriers in pathological alterations of the vascular system 
(such as venous or arterial thrombosis).
These predictions would depend on the original dimension of the  
spherical droplet along with the initial NP surface coverage, 
and the NP dimension to droplet size ratio, which would affect 
the surface area to volume ratio of the system and the average surface pressure of the particle–laden 
interface~\cite{2018-SoftMatter-Gu-Botto}, respectively.

To extend our results, the flow properties of the system are 
analyzed with two essential control parameters, \textit{i.e.} 
the Weber number ($We$) and the Ohnesorge number ($Oh$), 
commonly used in microfluidic~\cite{2013-CERD-Hall-Rothman,2020-PhysFluids-Xu-Che} and droplet formation \cite{Roas}. 
The Weber number, $We = \rho_o v^2 d_D/\gamma$, represents 
the ratio of the disrupting inertial force to the restorative 
surface tension force, where $\rho_o$ and $v$ are the density 
and the relative velocity of the ambient fluid (decane oil) 
and $d_D$ and $\gamma$ are the diameter and the interfacial 
tension of the droplet, respectively. 
The Ohnesorge number, $Oh = \mu_W /\sqrt{\rho_W \gamma d_D}$, 
represents the relative importance of the viscous force to the inertial 
and surface tension forces, where $\mu_W$ and $\rho_W$ are 
the dynamics viscosity and the density of the water droplet, respectively. 
From the calculation of $Oh$, one can define 
the critical Weber number, $We_C = 12~(1+1.5 \times Oh^{0.74})$, 
which corresponds to the minimum Weber number for a droplet 
to exhibit breakup modes~\cite{1996-PECS-Gelfand}.
Given $\gamma \sim 51.7~\textrm{mN.m}^{-1}$ the interfacial tension for a planar decane/water interface~\cite{2012-PRE-Fan-Striolo,2019-ACSNano-Sicard-Striolo}, 
$\rho_W \sim 1000~\textrm{kg.m}^{-3}$ and $\rho_o \sim 726~\textrm{kg.m}^{-3}$ the density of water and decane oil, respectively, 
$v \sim 50-70~\textrm{m.s}^{-1}$ the stationary 
velocity of the laminar flow (cf. Fig.\ref{fig3}a), 
$\mu_W = 8.9\times 10^{-4}~\textrm{Pa.s}$ the dynamics viscosity of 
water, and $d_D \sim 40$ nm the droplet diameter obtained from 
the measure of $R_{\textrm{GYR}}$ (cf. Fig.~\ref{fig2}a), 
we obtain $Oh \sim 0.6$, $We_C \sim 25$, and $We \sim 1.4 - 2.8$, 
indicating the armored droplets considered in the flow-asssisted 
encapslation and release processes are outside their breakup 
regime~\cite{2010-ARMR-Derby}.\\

Now, based on the estimation of the Weber number, 
we first extend our predictions to the high-shear regime of 
spreading water-in-oil/oil-in-water emulsion-based products. 
Given the relation $v \sim \dot{\epsilon}\times L_{\perp}$ with $L_{\perp}$ 
the dimension of the system orthogonal to the flow direction, 
we obtain $We \sim \rho_o \dot{\epsilon}^2 L_{\perp}^2 d_D/\gamma$.
Considering the average thickness of a cream $L_{\perp}\sim 1$ cm 
and representative shear rates $\dot{\epsilon} 
\sim 10^2 - 10^3 ~\textrm{s}^{-1}$~\cite{2019-IJAME-Walicka-Falicki,2020-Pharmaceutics-Simoes-Vitorino}, we obtain the characteristic 
dimension of the emulsion droplet 
$d_D \sim 1 - 100~\mu\textrm{m}$, corresponding 
to the minimal droplet size to observe the encapsulation or release mechanism, in agreement with the range of 
characteristic droplet sizes commonly used in topical pharmaceutical 
products~\cite{LU201059,2020-Pharmaceutics-Simoes-Vitorino}.

Either by skin adsorption or others intake paths, 
targeted carriers can reach bloodstream 
as required. The complexity of the flow scenarios 
present in the circulatory system defies the full description of the behaviour 
of our model carrier once entering into the body. 
However, it is possible to put our predictions  
into perspective with the transport of our model 
carrier in the vascular subsystem, in particular in the 
pathological flow conditions encountered in venous or 
arterial thrombosis~\cite{2009-BloodRev-Esmon}. 
The fluid properties of the hepatic artery in non-pathological 
conditions, which is representative of a large artery, 
has a characteristic dimension 
$L_{\perp}\sim 5 ~\textrm{mm}$, and shear rate  
$\dot{\epsilon} \sim 500~s^{-1}$~\cite{2015-FSOA-Sakariassen-Turitto}.
A pathological flow, on the other hand, can be defined 
as where the blood reaches shear rates 
$\dot{\epsilon} > 5000~s^{-1}$, resulting, for example, 
from pathological clotting of blood within the lumen of 
a vessel~\cite{2018-Biomicrofluidics-Herbig-Diamond}.
Considering $\rho_{\textrm{blood}}\sim 1060~\textrm{kg.m}^{-3}$ 
and $\gamma \sim 42~\textrm{mN.m}^{-1}$ as representative values 
of the average density and interfacial tension (against 
fluorocarbon) of the blood 
fluid~\cite{1981-JClinChemClinBiochem-Mottaghy-Hahn}, 
along with the narrowing of the pathological vessel 
$L_{\perp}\to L_{\perp}/2$, we obtain 
$d_D \sim 500~\textrm{nm}$ 
for the minimal droplet dimension in the 
conditions of the hepatic 
artery with pathological alterations to observe 
the encapsulation or release mechanism.
For comparison, we obtain  
$d_D \sim 10~\mu\textrm{m}$ for the minimal droplet 
dimension in the conditions of the normal hepatic artery,
in the range of sizes characteristic 
of leucocyte and red blood cells ~\cite{Phillips2009}. 
As a result, the process of targeted-delivery 
of active-compounds (such as antithrombotic agents) 
can be selectively controled with the size of the 
model nanocarrier.

\section*{Conclusions}
The numerical simulations discussed above allowed us to 
unravel the interplay between the structural morphology of 
armored nanodroplets and the organisation of the NP interfacial network, 
when the volume of the system is reduced, in qualitative agreement 
with experimental observation~\cite{2010-Langmuir-Datta-Weitz}. 
We showed that finite-size NPs can strongly affect the droplet 
shape with the formation of \textit{pocket-like} depressions, which 
can structurally favour the loading of a probe load. 
Eventually, our method would allow including specific interactions inside 
the formed cavity in order to mimic, for example, protein binding pockets 
or catalytic nanosurfaces.  

The dynamical response of specifically 
designed \textit{pocket-like} nanodroplets under different shear flow 
conditions exhibited the formation of long-lived 
anisotropic structures, characteristic of a jammed particle 
coverage at the liquid-liquid interface, associated with the 
dynamical rearrangement of the NP interfacial network.
Most importantly, the ability of \textit{pocket-like} 
nanodroplets to encapsulate or realease spherical solute loads, 
located inside the crater-like depression, during their transport under 
shear-flow conditions was validated.\\

Our predictions on the 
flow-assisted encapsulation and release of load probes 
in armored nanodroplets were extended  to systems in 
the micron scale encountered in pharmaceutical and 
cosmetic technology. 
Noticeably, we demonstrated that the mechanism 
reported in our work could be at play at larger scales, 
such as those encountered in the high-shear regime of 
spreading creams and ointments on the skin, and the 
transport of targeted carriers in 
pathological alterations of the vascular system.
We put the physical properties of our model carrier 
into perspective within the conditions encountered in 
the pathological alteration of the hepatic artery, where 
the formation of a blood clot inside the blood vessel can 
obstruct the flow of blood through the circulatory system 
increasing the haemodynamic shear stress and 
the risk of bleeding complications. 
In particular, hepatic artery thrombosis can be 
a very serious complication of liver transplantation, 
with mortality  in children which can be as high 
as $70\%$~\cite{ACHARYA2016279}. 
Hence, it is  essential to develop distinctive means to 
control the process of targeted-delivery of antithrombotic 
agents in the vascular system.\\

The physical insights discussed here provide a deeper understanding 
on the potential role played by nanoparticle-stabilized emulsions 
in the biomimetic design of novel hybrid materials for 
targeted-oriented active load delivery. This information could be useful for a variety of applications 
including the design of pharmaceutical carriers for drug delivery 
and pathogen encapsulation, where knowledge of the 
rheological properties of the system must be quantitatively assessed.

\section*{Acknowledgements}
F.S. acknowledges J. Reguera  for fruitful suggestions 
and A. Striolo for useful discussions. 
Via our membership of the UKs HEC Materials Chemistry Consortium,
which is funded by EPSRC (EP/L000202), this work
used the ARCHER UK National Supercomputing Service
(http://www.archer.ac.uk).

\normalem
\bibliographystyle{achemso.bst}
\bibliography{acs}

\section*{Methods}
\textbf{Mesoscopic framework.} The Dissipative Particle Dynamics (DPD) 
simulation method~\cite{Groot1997} is implemented within the simulation 
package LAMMPS~\cite{Plimpton1995}. 
In the DPD simulations, a particle represents a cluster of atoms rather 
than an individual atom. These particles interact with each other 
through soft particle-particle interactions. The movement of the 
particle can be realized by solving the Newton's equation of motion
\begin{equation}
    \frac{d\mathbf{r}_i}{dt}=\mathbf{v}_i \,, ~~~~~ m_i \frac{d\mathbf{v}_i}{dt} = \mathbf{F}_i\,,
\label{EqMotion}
\end{equation}
where $m_i$, $\mathbf{r}_i$, $\mathbf{v}_i$, and $\mathbf{F}_i$ denote 
the mass, position, velocity, and total force acting on the $i$th 
particle, respectively. The total force $\mathbf{F}_i$ is divided 
into three parts, the conservative force $\big(\mathbf{F}^C_{ij}\big)$, 
dissipative force $\big(\mathbf{F}^D_{ij}\big)$, and random 
force $\big(\mathbf{F}^R_{ij}\big)$, and defined as 
$\mathbf{F}_{i} = \sum_{j\neq i} \Big( \mathbf{F}^C_{ij} + \mathbf{F}^C_{ij}+\mathbf{F}^C_{ij}\Big)$ with
\begin{eqnarray}
\mathbf{F}^C_{ij} &=& a_{ij} \sqrt{\omega(r_{ij})}\,\hat{\mathbf{r}}_{ij} \,, \label{Fcons}\\
\mathbf{F}^D_{ij} &=& -\Gamma \omega(r_{ij}) (\hat{\mathbf{r}}_{ij} \cdot \mathbf{v}_{ij})\hat{\mathbf{r}}_{ij} \,,\\
\mathbf{F}^R_{ij} &=& \sigma \sqrt{\omega(r_{ij})} \theta_{ij}\hat{\mathbf{r}}_{ij} \,
\end{eqnarray}
where $\mathbf{r}_{ij} = \mathbf{r}_{i}-\mathbf{r}_{j}$, 
$r_{ij} = \vert \mathbf{r}_{ij} \vert$, 
$\hat{\mathbf{r}}_{ij} = \mathbf{r}_{ij}/r_{ij}$, and 
$\mathbf{v}_{ij} = \mathbf{v}_{i}-\mathbf{v}_{j}$. 
The weight function $\omega(r_{ij})$ equals to $(1-r_{ij}/R_c)^2$ with 
a cut-off distance $R_c$. $a_{ij}$, $\Gamma$, $\sigma$, and $\theta_{ij}$ 
are the repulsive parameter, friction coefficient, noise amplitude, 
and Gaussian random variable, respectively. To keep the temperature 
of the system constant, $\Gamma$ and $\sigma$ satisfy the 
fluctuation-dissipation theorem as $\sigma^2=2\Gamma k_B T$, where 
$k_B$ and $T$ are the Boltzmann and the absolute temperature, respectively.

The system simulated here is composed of water, oil (decane), 
nanoparticles (NPs) and solute molecules. Following previous work~\cite{2013-Langmuir-Luu-Striolo,2016-FD-Sicard-Striolo,2017-Nanoscale-Sicard-Striolo,2019-ACSNano-Sicard-Striolo}, 
we choose the degree of coarse graining $N_m = 5$ with the understanding 
that one "water bead" (w) represents $5$ water molecules. 
Within this assumption, the volume of each bead is 
$V_{\textrm{bead}} \approx 150 {\AA}^3$.
The scaled density is set to $\rho = 3$ beads/$R_c^3$, 
where $R_c$ is the DPD cutoff distance given as 
$R_c = \sqrt[3]{\rho V_{\textrm{bead}}} \approx 0.766$ nm.
The scaled mass of each bead (oil, water, solute molecule, 
and NP beads) was set to 1. One decane molecule is modeled 
as two "oil beads" (o) connected by one harmonic spring of 
length $0.72$ $R_c$ and spring constant $350$ $k_BT/R_c$~\cite{Groot2001}. 
The size of the triclinic simulation box (initially orthogonal) 
is $L_x \times L_y \times L_z \equiv 200 \times 100 \times 100$ $R_c^3$, 
where $L_x$ (respectively $L_y$ and $L_z$) is the box length along 
the $X$ (respectively $Y$ and $Z$) direction. 
Periodic boundary conditions are applied in all three directions.
The solute molecules and the NPs are modelled as hollow rigid 
spheres, as already described in previous work~\cite{2013-Langmuir-Luu-Striolo,2016-FD-Sicard-Striolo,2017-Nanoscale-Sicard-Striolo,2019-ACSNano-Sicard-Striolo}.
The hydrophobic solute molecules are made of nonpolar DPD beads, 
whereas the NPs contain polar (p) and nonpolar (ap) DPD beads 
on their surface~\cite{Calvaresi2009}. 
One DPD bead was placed at the NP and solute molecule centers 
for convenience, as described elsewhere~\cite{2013-Langmuir-Luu-Striolo, LuuJPCB2013}. 
All types of beads in our simulations have reduced mass of $1$. 
We maintain the surface bead density on the NPs and solute molecule 
sufficiently high to prevent other DPD beads (either decane or water) 
from penetrating the NPs and solute molecules~\cite{LuuJPCB2013}.\\

The interaction parameters shown in Table \ref{Tab-interaction} are used here. 
These parameters are adjusted to reproduce selected atomistic simulation 
results, as explained in prior work~\cite{2013-Langmuir-Luu-Striolo}. 
The interaction parameters between NP polar and nonpolar beads, 
as well as solute molecule beads, are adjusted to ensure 
that NPs/NPs and NPs/solute are able to assemble and disassemble 
without yielding permanent dimers at the water/oil 
interface~\cite{2013-Langmuir-Luu-Striolo}.
The scaled temperature was set to $1$, equivalent to $298.73$ K. 
The time step $\delta t = 0.03 \times \tau$ was used to integrate the equations 
of motion, where $\tau$ is the DPD time constant. As demonstrated by 
Groot and Rabone~\cite{Groot2001}, the time constant of the simulation can
be gauged by matching the simulated self-diffusion of water, $D_{\textrm{sim}}$, 
with the the experimental water self-diffusion coefficient, 
$D_{\textrm{water}}= 2.43 \times 10^{-5}$ $\textrm{cm}^2/s$~\cite{Partington1952}, 
calculated as $\tau = \frac{N_m D_{\textrm{sim}} R_c^2}{D_{\textrm{water}}}$, 
as shown in previous work~\cite{2013-Langmuir-Luu-Striolo}.
When $a_{w-w} = 131.5$ $k_B T/R_c$, this results in a time step 
$\delta t = 5.6$ ps.\\ 
\begin{table}[h]
\begin{center}
\begin{tabular*}{0.45\textwidth}{@{\extracolsep{\fill}}lccccc}
  \hline
  {} & $w$ & $o$ & $ap$ & $p$ & $s$\\
  \hline
  $w$ & $131.5$ & $198.5$ & $178.5$ & $110$ & $670$\\
  $o$ & {} & $131.5$ & $161.5$ & $218.5$ & $161.5$\\
  $ap$ & {} & {} & $450$ & $670$ & $450$\\
  $p$ & {} & {} & {} & $450$ & $670$\\
  $s$ & {} & {} & {} & {} & $131.5$\\
  \hline
\end{tabular*}
\caption{DPD interaction parameters expressed in $k_BT/R_c$ units. 
Symbols $w$, $o$, $ap$, $p$, and $s$ stand for water beads, oil beads, 
NP nonpolar beads, NP polar beads, and solute beads, respectively.}
\label{Tab-interaction}
\end{center}
\end{table}

While the DPD framework satisfies the Navier-Stokes equations in the continuum 
limit~\cite{Groot1997}, the traditional DPD algorithm cannot reproduce 
the vapour-liquid coexistence of water at the droplet interface~\cite{2003-PRE-Warren}. 
This is due to the DPD conservative force, which determines the 
thermodynamics of the DPD system and yields the equation of state (EOS)~\cite{Groot1997}
\begin{equation}
p = \rho k_B T + \alpha a \rho^2 ~,
\label{EOS}
\end{equation}
where $p$ is the pressure, $\rho$ is the number density of the DPD beads, 
$a$ is the repulsion strength, and $\alpha$ is a fitting parameter equal 
to $0.101 \pm 0.001$ in DPD reduced units~\cite{Groot1997}.
As shown by Warren in Ref.~\cite{2003-PRE-Warren}, The DPD system is unstable 
for $a<0$, so one is restrictided to $a \geq 0$ and therefore to strictly 
repulsive (conservative) interactions. This implies that calculations 
such as the vapor-liquid coexistence and free-surface simulations 
cannot be attempted. This can be adjusted by considering higher order terms 
of the density, $\rho$, in Eq.~(\ref{EOS}), \textit{i.e.} making the conservative 
force in Eq.~(\ref{Fcons}) density dependent~\cite{2003-PRE-Warren}.\\

\textbf{Nonequilibrium simulation.} To simulate the response of the 
 system subjected to an homogeneous shear flow, we employ the SLLOD 
 algorithm~\cite{1984-CPR-Evans-Morriss,1984-PRA-EvansMorris} coupled 
 with the Lee-Edwards periodic boundary conditions~\cite{1972-JPC-Lees-Edwards}, 
 as implemented in the simulation package LAMMPS~\cite{Plimpton1995}. 
The SLLOD algorithm modifies the equations of motion in Eq.~\ref{EqMotion} as:
\begin{eqnarray}
\frac{d\mathbf{r}_i}{dt} &=& \mathbf{v}_i + \mathbf{e}_x \dot{\epsilon} r_{i,y} \,, \\
m_i\frac{d\mathbf{v}_i}{dt} &=& \mathbf{F}_i - m_i \mathbf{e}_x \dot{\epsilon} v_{i,y} \,,
\end{eqnarray}
where $\dot{\epsilon} = \partial v_x/ \partial r_y$ is the shear rate 
of the external flow and $\mathbf{e}_{x,y}$ are the unit vectors 
along the $x$ and $y$ directions, respectively. 
The velocity of the $i$th particle is divided into two parts, that is, 
the peculiar velocity $\mathbf{v}_i$ representing the random thermal motions 
and the shear flow velocity $\mathbf{e}_x \dot{\epsilon} v_{i,y}$ 
relating to the external disturbance strength. 
Specifically, we impose a linear velocity profile in the $x$ direction 
with a constant gradient in the $y$ direction, keeping the density of 
the system constant, by changing the $xy$-tilt factor, $T_{xy}$, of the 
triclinic simulation box at a constant shear rate, 
$\dot{\epsilon}$, as 
\begin{equation}
    T_{xy}(t) = T_{xy}^{(0)} + \dot{\epsilon}~L_0~\Delta t\,.
\label{ShearEq}
\end{equation}
In Eq.~\ref{ShearEq}, $T_{xy}^{(0)}$ and $L_0$ are the initial 
tilt factor and the original length of the box perpendicular 
to the shear direction. This can be related to the shear stress 
of the external shear flow $\tau_s = \mu \dot{\epsilon}$, with $\mu$ 
the dynamic viscosity of the continuous phase.\\

\textbf{Three phase contact angle.} To estimate the three phase 
contact angle, $\theta_C$, for the NPs on the droplets we calculate 
the fraction of the spherical NP surface 
area that is wetted by water~\cite{FanSM2012},
\begin{equation}
 \theta_C = 180 - \arccos\Big(1-\frac{2 A_w}{4\pi R^2}\Big) \,,
\end{equation}
where $A_w$ is the area of the NP surface that is wetted 
by water and $R$ is the radius of the NP. The ratio $A_w/4\pi R^2$ 
is obtained by dividing the number of NP surface beads (ap or p), 
which are wetted by water, by the total number of beads on the NP 
surface. One surface bead is wet by water if a water bead is 
the solvent bead nearest to it. One standard deviation from 
the average is used to estimate the statistical uncertainty.\\

\textbf{Interfacial tension.}The interfacial tension $\gamma$ at 
the water/oil interface as a function of the NP surface coverage 
$\Phi$ is calculated as~\cite{2012-PRE-Fan-Striolo,2013-Langmuir-Luu-Striolo}
\begin{equation}
\gamma = \Bigg\langle P_{zz}-\frac{P_{xx}+P_{yy}}{2} \Bigg\rangle \frac{L_z}{2} \,.
\label{InterfTension}
\end{equation}
In Eq.~\ref{InterfTension}, $P_{ij}$ is the $ij$ element of 
the pressure tensor, $L_z$ is the simulation box length 
in the $z$ dimension, and the angular brackets denote the 
ensemble average.\\

\textbf{Self-diffusion coefficient.} To characterize the self-diffusion 
coefficient of the NPs at the water/oil interface, we estimate 
the mean squared displacement (MSD) of a single NP adsorbed at a planar 
interface parallel to the $x-y$ plane. For each particle size, 
the simulated diffusion coefficient is estimated according to
\begin{equation}
    D_{x-y}=\frac{1}{4} \Bigg\langle \frac{\vert r_i(t)-r_i(0)\vert^2}{t}\Bigg\rangle
\end{equation}
where $r_i(t)$ is the position of particle $i$ at time $t$ on the 
plane of the interface.\\

\textbf{Gyration tensor.} To measure the evolution of the structural 
morphology of the emulsion droplet, we estimate the principal components 
of the gyration tensor~\cite{2011-JPCA-Vymetal-Vondrasek,Solc1971,2016-FD-Sicard-Striolo}, 
which allow the evaluation of the overall shape of the system 
and reveal its symmetry. Considering the definition for the gyration 
tensor, 
\begin{equation}
\mathcal{T}_{GYR} = \frac{1}{N}
\begin{bmatrix}
\sum x_i^2 & \sum x_i y_i & \sum x_i z_i \\
\sum x_i y_i & \sum y_i^2 & \sum y_i z_i \\
\sum x_i z_i & \sum y_i z_i & \sum z_i^2
\end{bmatrix} \, ,
\end{equation}
where the summation is performed over $N$ atoms and the coordinates 
$x$, $y$, and $z$ are related to the geometrical center of the atoms, 
one can define a reference frame where $\mathcal{T}_{GYR}$ can be 
diagonalized:
\begin{equation}
\mathcal{T}_{GYR}^{diag} =
\begin{bmatrix}
S_1^2 & 0 & 0\\
0 & S_2^2 & 0 \\
0 & 0 & S_3^2
\end{bmatrix} \, .
\label{Tgyr}
\end{equation}
In Eq.~\ref{Tgyr}, we follow the convention of indexing the eigenvalues 
according to their magnitude, \textit{i.e.} $S_1^2 > S_2^2 > S_3^2$. 
We define the radius of gyration $R_{GYR}^2 \equiv S_1^2 + S_2^2 + S_3^2$ 
and the relative shape anisotropy 
$\kappa^2 = \frac{3}{2} \frac{S_1^4 + S_2^4 + S_3^4}{(S_1^2 + S_2^2 + S_3^2)^2} - \frac{1}{2}$, 
and we calculate $R_{GYR}$ and $\kappa^2$ using the centers of the water beads.

\pagebreak
\widetext
\begin{center}
\textbf{\large Computational design of armored nanodroplets as nanocarriers for encapsulation and release under flow conditions} \end{center}

\begin{center}\textbf{\large Supporting Information}
\end{center}
\setcounter{equation}{0}
\setcounter{figure}{0}
\setcounter{table}{0}
\setcounter{page}{1}
\makeatletter
\renewcommand{\theequation}{S\arabic{equation}}
\renewcommand{\thefigure}{S\arabic{figure}}

\vskip 1.0cm
\section*{Nanoparticle characteristics}
\begin{figure}[t]
\includegraphics[width=0.9 \textwidth, angle=-0]{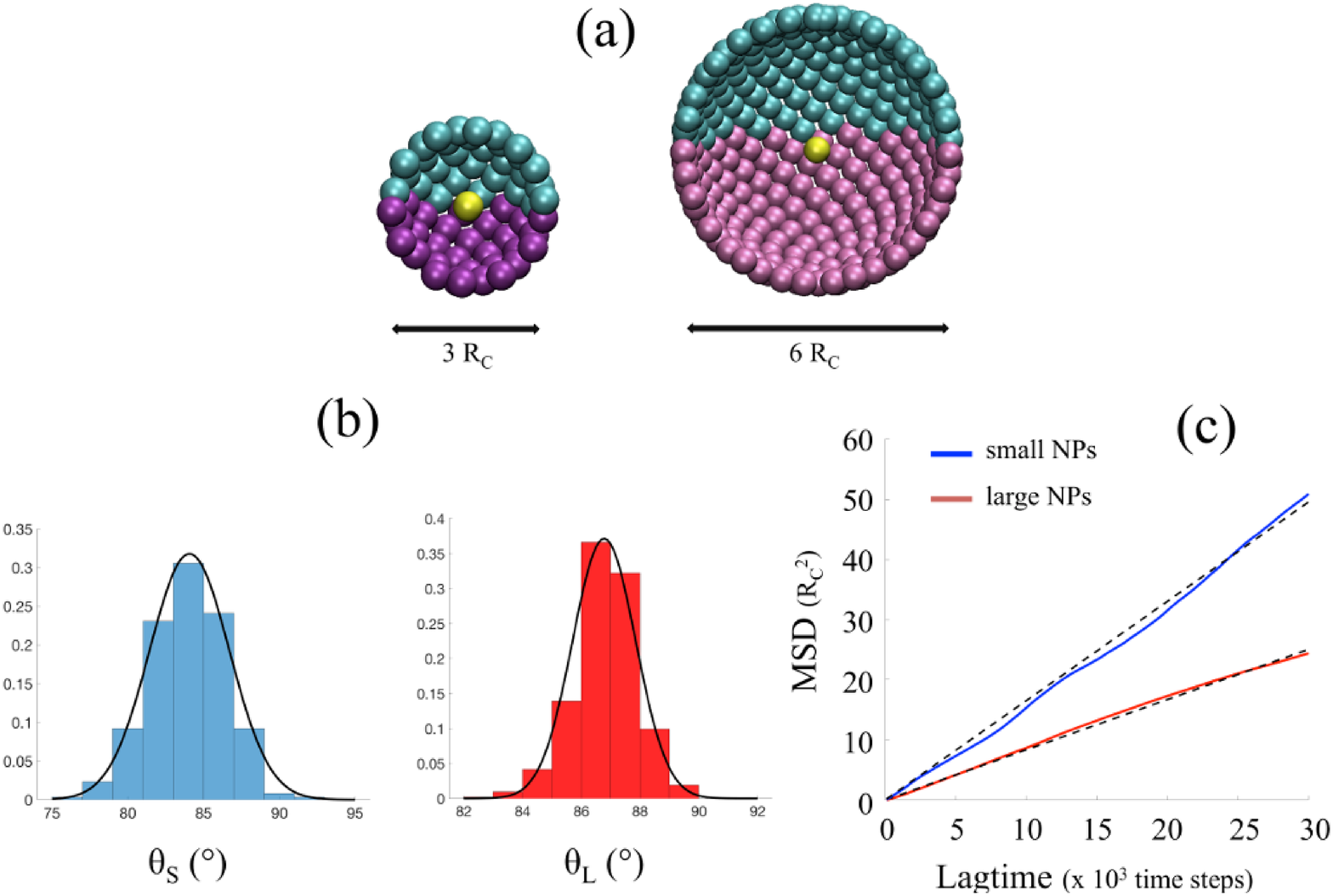}
 \caption{(a) Cross sectional view of the small (left panel) and large 
 (right panel) spherical NPs simulated in this work. 
 Cyan, purple and gold spheres represent the nonpolar (ap), polar (p), 
 and NP center beads, respectively. Small and large NPs are covered 
 with 108 and 432 beads,respectively, corresponding to a surface density 
 of $\sim 3.8$ beads per $R_c^2$ on the NP surface. 
 The fractions of nonpolar and polar beads on the NP surface are 
 identical. 
 (b) Probability distributions of the three-phase contact angles 
 $\theta_S$ and $\theta_L$ for small (S) and large (L) NPs, respectively. 
 The probability distributions is fitted with Gaussian distributions 
 of means $\mu_S \sim 84.1^{\circ}$ and $\mu_L \sim 86.8^{\circ}$, 
 and standard deviations $\sigma_S \sim 2.7^{\circ}$ and $\sigma_L \sim 1.1^{\circ}$, 
 as shown with continuous lines.
 (c) MSD as a function of simulation lagtime for small and large NPs 
 measured at the water/oil planar interface.}
\label{figS1}
\end{figure}

Following previous work~\cite{2011-Langmuir-Fan-Striolo,2012-PRE-Fan-Striolo,2013-Langmuir-Luu-Striolo,2016-FD-Sicard-Striolo,2017-Nanoscale-Sicard-Striolo,2019-ACSNano-Sicard-Striolo},
the nanoparticles (NPs) are specifically designed to represent 
Janus silica NPs (particles whose surface shows two distinct 
wetting properties) at the decane/water interface.
The NPs are modelled as hollow rigid spheres with two different diameters, 
$d_S \sim 3 R_c$ and $d_L \sim 6 R_c$ for small and large NPs, respectively, 
with $R_c \sim 0.766$ nm the DPD cutoff distance.
Each NP contains polar (p) and nonpolar (ap) DPD beads on its surface and 
one DPD bead is placed at the NP center for convenience, as shown in Fig.~\ref{figS1}a.
Hollow models have been used in the literature to simulate NPs, 
and hollow NPs can also be synthesized experimentally~\cite{Calvaresi2009}. 
All types of beads in our simulations have reduced mass of $1$. 
To cover small and large NPs, $108$ and $432$ beads are required, 
respectively, yielding a surface density of $\approx 3.8$ beads per $R_c^2$ 
on the NP surface~\cite{2012-PRE-Fan-Striolo}. 
The total number of beads on one NP surface is chosen such that 
the surface bead density be sufficiently high to prevent other DPD beads 
(either decane or water) from penetrating the NPs (which would be unphysical), 
as it has already been explained elsewhere~\cite{LuuJPCB2013}.
We use the same surface density for the hydrophobic spherical probe load.\\

The NP-solvent interaction parameters in the DPD framework, 
given in the Methods section in the main text, 
were originally parametrized to reproduce the three-phase contact angle, 
$\theta_c \sim 85.3^{\circ} \pm 1.9^{\circ}$,  
obtained via atomistic molecular dynamics (MD) simulations 
for one silica Janus NP of diameter $\sim 2 R_c$ 
at the decane/water interface, as explained in previous work~\cite{2011-Langmuir-Fan-Striolo,2012-PRE-Fan-Striolo,2013-Langmuir-Luu-Striolo}. 
In our case, we check that the three-phase contact angles for small 
and large NPs, $\theta_S \sim 84.1^{\circ} \pm 2.7^{\circ}$ 
and $\theta_L \sim 86.8^{\circ} \pm 1.1^{\circ}$, respectively, 
as shown in Fig.~\ref{figS1}b, are in qualitative agreement, 
within the standard errors, with experimental 
observations~\cite{2010-PCCP-Arnaudov-Paunov}. 
From the error bars measured, we observe that the small NPs are
more sensitive to thermal fluctuations at the interface
compared to the large ones, characteristic of the increase
of the adsorption energy with the particle radius~\cite{2001-Langmuir-Binks-Fletcher,2007-JCP-Jiang-Granick,2020-PCCP-Khedr-Striolo}.\\

To evaluate the diffusion of the small and large NPs at the water/oil 
interface, we estimate the mean squared displacement (MSD) of a single 
NP adsorbed at a planar water/oil interface parallell to the $x-y$ plane 
for increasing simulation lagtime, as shown in Fig.~\ref{figS1}c. 
For each particle size, the MSD is averaged over $5$ replicas conducted 
for $1~\mu \textrm{s}$ each, and the simulated diffusion coefficient 
is estimated accordingly (see Methods section in the main text). 
We measure
$D_S \sim 4.7 \pm 3.1 \times 10^{-7}$ $\textrm{cm}^2~\textrm{s}^{-1}$ 
and 
$D_L \sim 1.8 \pm 0.7 \times 10^{-7}$ $\textrm{cm}^2~\textrm{s}^{-1}$, 
for small and large NPs, respectively. 
In particular, large NPs are less diffusive than smaller ones, 
in qualitative agreement with simulations~\cite{2020-PCCP-Khedr-Striolo} 
and experimental observations~\cite{2011-Small-Wang-Koynov}.
\newpage

\section*{Formation of pocket-like structures}
The number of water beads constituting the initial water-in-oil emulsion 
droplets is fixed to $N_W \approx 3\times 10^5$. At the beginning of 
each simulation, the solvent (oil) beads are uniformly distributed within 
the simulation box. One water droplet of radius $\approx 32~R_C$ is generated 
by replacing the oil beads with water beads within the volume of the spherical 
surface. A number of spherical NPs are placed randomly at the water-decane 
interface with their polar (nonpolar) part in the water (oil) phase to achieve 
the desired water-decane interfacial area per NP. 
The initial configuration obtained is simulated for $10^6$ timesteps in order 
to relax the density of the system and the contact angle of the NPs 
on the droplet. The system pressure and the three-phase contact angle 
distributions converged after 5000 simulation steps.
Then, we let the system run for an additional $2\times 10^6$ timesteps 
to generate two new initial configurations, which allows us to test 
the reproducibility of the simulations.\\

To study the surface mechanical instabilities and the collapse 
mechanisms responsible for the formation of the crater-like depressions 
at the droplet interface, we follow the numerical protocol discussed 
by Sicard et al. in previous work~\cite{2017-Nanoscale-Sicard-Striolo}. 
The surface area of the droplets is slowly diminished, pumping 
randomly a constant proportion, \textit{i.e.} $10$ percent, of water 
molecules out of the droplet and letting the system pressure and 
the three-phase contact angle distribution equilibrate at constant density. 
By slowly, we mean we do not create any hollow volume 
in the droplet that would strongly drive the system out-of-equilibrium. 
Doing so, the three-phase contact angle distribution of the NPs 
evolves sufficiently smoothly when the droplet buckles and becomes 
nonspherical, thereby preventing particles to be artifactually realeased. 
This numerical protocol is comparable to a solubilization experiment, where 
the dispersed phase is slightly soluble in the continuous 
phase~\cite{2010-Langmuir-Datta-Weitz}. 
By adding a fixed amount of unsatured continuous phase, the volume of the droplets 
can then be controllably reduced.\\

\begin{figure}[b]
\includegraphics[width=0.75 \textwidth, angle=-0]{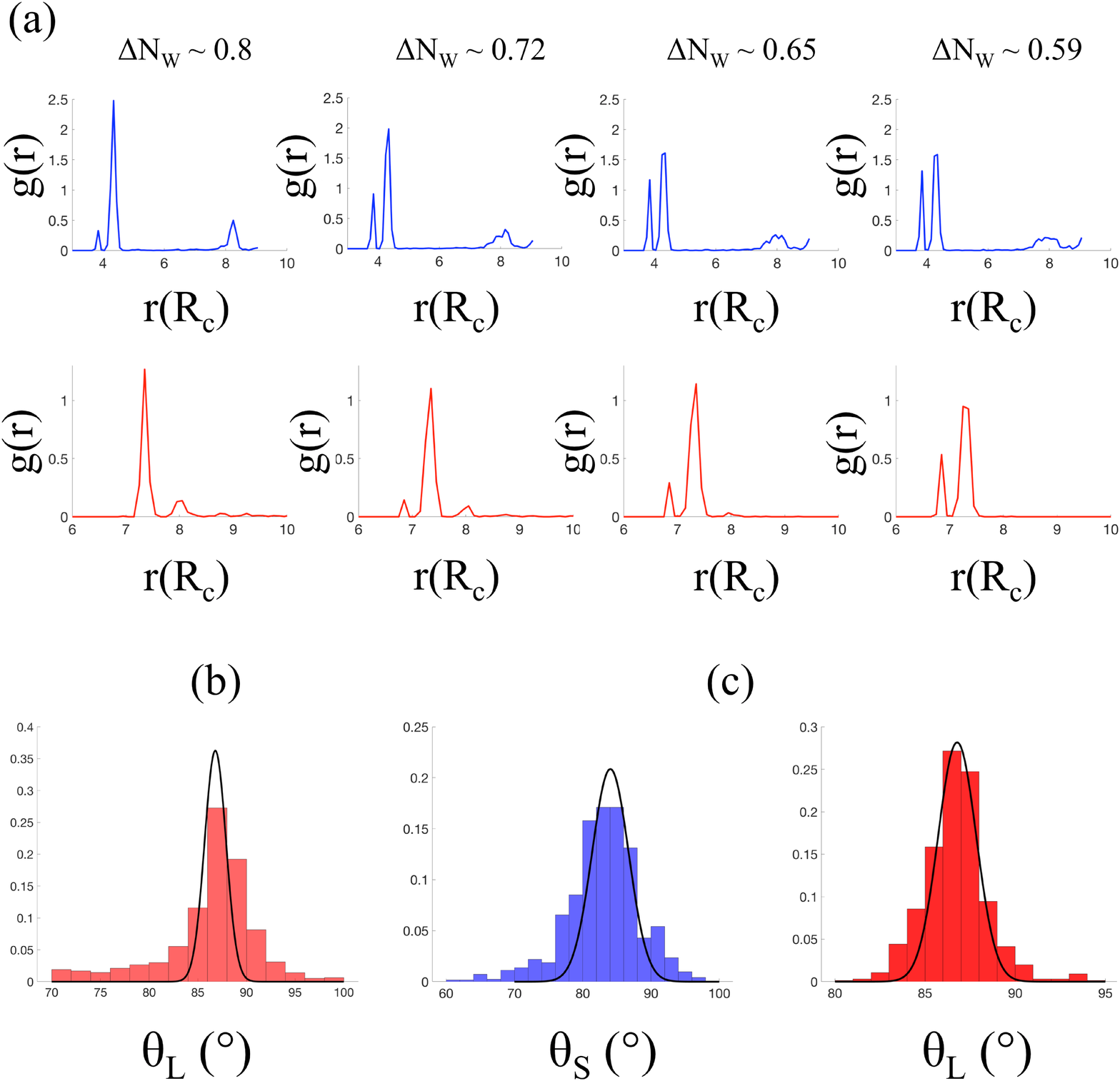}
 \caption{(a) Evolution of the NP radial distribution function, $g(r)$, 
 as a function of the dimensionless parameter $\Delta N_W$, defined in the 
 main text, when the droplet is uniformly covered with small (top panel) 
 and large (bottom panel) NPs.
 (b) Probability distribution of the three-phase contact angle 
 of large NPs, $\theta_L$, at the interface of $\textrm{UC}_L$, 
 when $\Delta N_W \sim 0.35$. The initial Gaussian distribution, 
 fitted with continuous line, is shown for comparison.
(c) Probability distributions of the three-phase contact angle of 
$\theta_S$ and $\theta_L$, for small and large NPs, respectively, 
at the interface of $\textrm{HC}_1$, when $\Delta N_W \sim 0.35$. 
The initial Gaussian distributions, fitted with continuous lines, 
are shown for comparison.
 }
\label{figS2}
\end{figure}

To study quantitatively the transition from spherical shrinking 
to buckling in the uniformly covered droplets, 
$\textrm{UC}_S$ and $\textrm{UC}_L$, we follow the evolution 
of the radial distribution functions, $g_S(r)$ and 
$g_L(r)$, with $r$ the distance between the center of the NPs, 
along with the distributions of the three-phase contact angles, 
$\theta_S$ and $\theta_L$, of small and large NPs, respectively.
In Fig.~\ref{figS2}a, we show the evolution of $g(r)$, as a function of the 
dimensionless parameter $\Delta N_W$ defined in the main text, 
for $\textrm{UC}_S$ (blue) and $\textrm{UC}_L$ (red). Unlike $\textrm{UC}_S$ 
where the first peak in $g(r)$ is already present for $\Delta N_W \sim 0.8$ 
and increases significantly when $\Delta N_W$ decreases, we observe 
the apparition of the first peak in $g(r)$ for $\textrm{UC}_L$ at a later 
stage ($\Delta N_W \sim 0.72$). This peak increases significantly slower when 
$\Delta N_W$ decreases. This behaviour is representative of the difference 
in NP interfacial packing as a function of the NP size, with a transition 
from spherical shrinking to buckling happening when the NP monolayer 
becomes close to its maximum packing. 
When the volume of $\textrm{UC}_S$ and $\textrm{UC}_L$ is reduced, 
$\theta_C^{(S)}$ and $\theta_C^{(L)}$ uniformly evolve from a Gaussian 
to a skewed unimodal distribution, as shown in Fig.~\ref{figS2}b 
for $\textrm{UC}_L$, in line with previous work~\cite{2017-Nanoscale-Sicard-Striolo}. 
When the volume of $\textrm{HC}_1$ or $\textrm{HC}_2$ is reduced, 
on the other hand, we observe significant differences in the 
evolution of the distributions of $\theta_C^{(S)}$ 
and $\theta_C^{(L)}$, due to heterogeneity in NP size and 
surface coverage, as shown in Fig.~\ref{figS2}c.
In particular, the distribution of $\theta_C^{(L)}$ remains 
similar to the Gaussian distribution observed in the initial configuration 
(continuous line), while the distributions of $\theta_C^{(S)}$ shows 
larger variability, characterized with the increase of the asymmetry of 
the distribution towards lower values of $\theta_S$.

\newpage

\section*{Evolution of the structural morphology of the droplets under flow conditions}
\begin{figure}[h]
\includegraphics[width=0.8 \textwidth, angle=-0]{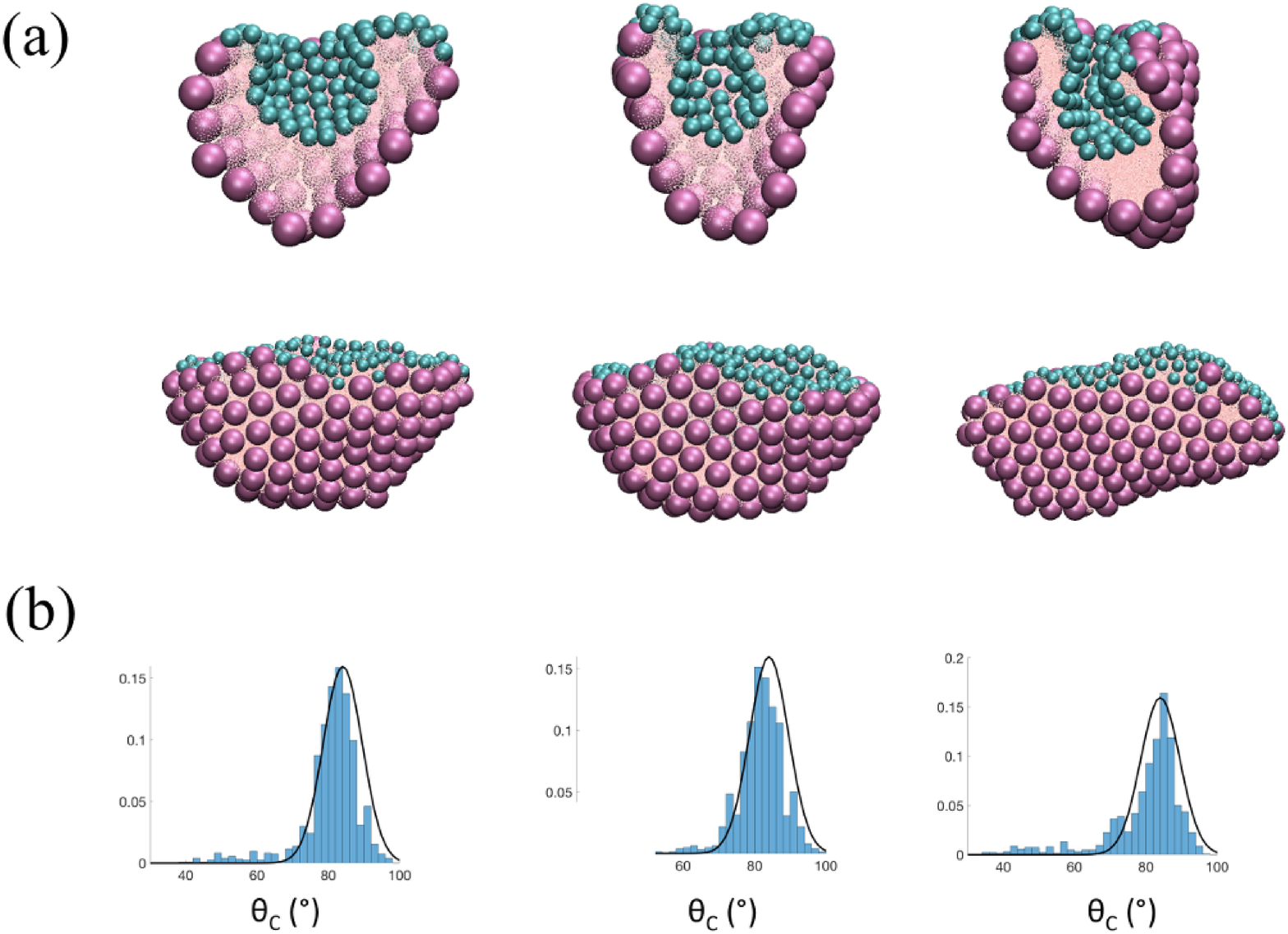}
 \caption{
 (a) Representative cross-view (top) and side-view (bottom) 
 of $\textrm{HC}1_{a,b,c}$ (from left to right) obtained after 
 the relaxation of the system ($t \sim 1.2~\mu s$). 
 Cyan and purple spheres represent the small and large Janus NPs, 
 respectively. Pink spheres represent water beads. 
 The oil molecules surrounding the system are not shown for clarity. 
 (b) Corresponding distributions of the three-phase contact angle, 
 $\theta_C^{(S)}$, for small NPs, at the interface of the structures 
 $\textrm{HC}1_{a,b,c}$.
 }
\label{figS3}
\end{figure}

As explained in details in the main text and the Methods section, 
we investigate the dynamical response of the buckled armored 
nanodroplets $\textrm{HC}_1$ subjected to shear flow of the surrounding 
fluid, using the SLLOD algorithm~\cite{1984-CPR-Evans-Morriss,1984-PRA-EvansMorris} 
coupled with Lee-Edwards periodic boundary conditions~\cite{1972-JPC-Lees-Edwards}.
The changes in the structural morphology of the system are characterized 
with the elongation of the nanodroplet along the deformation axis $x$, 
and the squeezing of the crater-like depression along the orthogonal 
$z$-direction, as shown in Fig.~\ref{figS3}a.
In Fig.~\ref{figS3}b, we show the probability distribution of 
the three-phase contact angle, $\theta_C^{(S)}$, for small NPs, 
at the interface of the structures $\textrm{HC}1_{a,b,c}$ defined 
in the main text.
Within the range of shear rates considered in this work, 
$\theta_C^{(S)}$ shows a  skewed unimodal distribution with a central  peak located at  the  same  value  
as  the one  measured  for  both the  initial and buckled 
configurations (shown with continuous lines).\\

\end{document}